\documentclass[a4paper,11pt]{article}
\pdfoutput=1 

\usepackage{jinstpub} 
\usepackage{comment}
\usepackage{graphicx}
\usepackage{subfig}

\usepackage[]{siunitx}  

\usepackage{upgreek}

\title{\boldmath Reflectivity and PDE of VUV4 Hamamatsu SiPMs in Liquid Xenon}

\author[a]{P.~Nakarmi}
\author[a,1]{{I.~Ostrovskiy}\note{Corresponding author}}
\author[a]{A.K.~Soma}
\author[b]{F.~Reti\`{e}re}
\author[c]{S.~Al Kharusi}
\author[d]{M.~Alfaris}
\author[e]{G.~Anton}
\author[f]{I.J.~Arnquist}
\author[g, 2]{{I.~Badhrees}
\note{also at King Abdulaziz City for Science and Technology, Riyadh, Saudi Arabia}}
\author[h]{P.S.~Barbeau}
% ...
\affiliation[a]{Department of Physics and Astronomy, University of Alabama, Tuscaloosa, AL 35487, USA}
\affiliation[b]{TRIUMF, Vancouver, British Columbia V6T 2A3, Canada}
\affiliation[c]{Physics Department, McGill University, Montr\'eal, Qu\'ebec H3A 2T8, Canada}
\affiliation[d]{Amherst Center for Fundamental Interactions and Physics Department, University of Massachusetts, Amherst, MA 01003, USA}
\affiliation[e]{Erlangen Centre for Astroparticle Physics (ECAP), Friedrich-Alexander University Erlangen-N\"urnberg, Erlangen 91058, Germany}
\affiliation[f]{Pacific Northwest National Laboratory, Richland, WA 99352, USA}
\affiliation[g]{Department of Physics, Carleton University, Ottawa, Ontario K1S 5B6, Canada}
\affiliation[h]{Department of Physics, Duke University, and Triangle Universities Nuclear Laboratory (TUNL), Durham, NC 27708, USA}
% ...

\emailAdd{iostrovskiy@ua.edu}

\abstract{Understanding reflective properties of materials and photodetection efficiency (PDE) of photodetectors is important for optimizing energy resolution and sensitivity of the next generation neutrinoless double beta decay, direct detection dark matter, and neutrino oscillation experiments that will use noble liquid gases, such as nEXO, DARWIN, DarkSide-20k, and DUNE. Little information is currently available about reflectivity and PDE in liquid noble gases, because such measurements are difficult to conduct in a cryogenic environment and at short enough wavelengths. Here we report a measurement of specular reflectivity and relative PDE of Hamamatsu VUV4 silicon photomultipliers (SiPMs) with 50 $\mu$m micro-cells conducted with xenon scintillation light ($\sim$175 nm) in liquid xenon. The specular reflectivity at 15$^{\circ}$ incidence of three samples of VUV4 SiPMs is found to be 30.4$\pm$1.4\%, 28.6$\pm$1.3\%, and 28.0$\pm$1.3\%, respectively. The PDE at normal incidence differs by $\pm$8\% (standard deviation) among the three devices. The angular dependence of the reflectivity and PDE was also measured for one of the SiPMs. Both the reflectivity and PDE decrease as the angle of incidence increases. This is the first measurement of an angular dependence of PDE and reflectivity of a SiPM in liquid xenon. 
}


\begin{document}
\maketitle
\flushbottom

\section{Introduction}
\label{sec:intro}
Understanding light collection in a large liquid xenon (LXe) detector is important for optimizing energy resolution and sensitivity of the next generation neutrinoless double beta decay~\cite{nEXO} and direct detection dark matter experiments~\cite{Darwin}. The task of modeling and optimizing the light collection and detection is significantly complicated
by the fact that LXe scintillates in the vacuum ultraviolet (VUV) region. Little information is currently available about reflective properties of materials and photodetectors in this region. Even when the data are available for short enough wavelengths, they are usually collected in vacuum~\cite{pavia_2007,Newell:97} or gas~\cite{solovov:2009,solovov:2010}, due to the difficulty of performing measurements in a cryogenic liquid. Extrapolating to the case of LXe may not be straightforward. Apart from the difference in the refraction indices in vacuum and LXe, the LXe may be altering the optical properties of some materials, especially the ones with porous surfaces~\cite{solovov:2009}. For example, it was found~\cite{lz:2015, Neves_2017} that the reflectance of Teflon (PTFE) films in LXe is not well predicted by the optical models derived from the vacuum measurements. The situation with reflectivity of silicon photomultipliers (SiPMs) is further complicated. Extrapolating the reflectivity~\footnote{We prefer to use the term ``reflectivity'' here, instead of ``reflectance'', due to the lack of information about the surface of some SiPMs} measured in vacuum to LXe requires knowledge of the composition and thickness of the thin top layers, which is often unavailable. The above challenges are not unique to LXe. Understanding light collection and reflectivity in the VUV is similarly important to maximize the sensitivity of future experiments that use liquid argon, e.g., DarkSide-20k~\cite{DS20k} and DUNE~\cite{dune:2016}.

nEXO is a planned 5-tonne LXe detector that will look for neutrinoless double beta decay of $^{136}$Xe~\cite{nexo_pcdr}. nEXO plans to use SiPMs to detect the xenon scintillation light.
SiPMs produced by FBK~\footnote{Fondazione Bruno Kessler, https://www.fbk.eu.} and Hamamatsu~\footnote{Hamamatsu Photonics, https://www.hamamatsu.com} are currently being considered by nEXO. To make an optimal choice of SiPMs, their reflectivity and photodetection efficiency (PDE) in LXe need to be understood.

\begin{figure}[h]
 \centering
\includegraphics[width=0.45\textwidth]{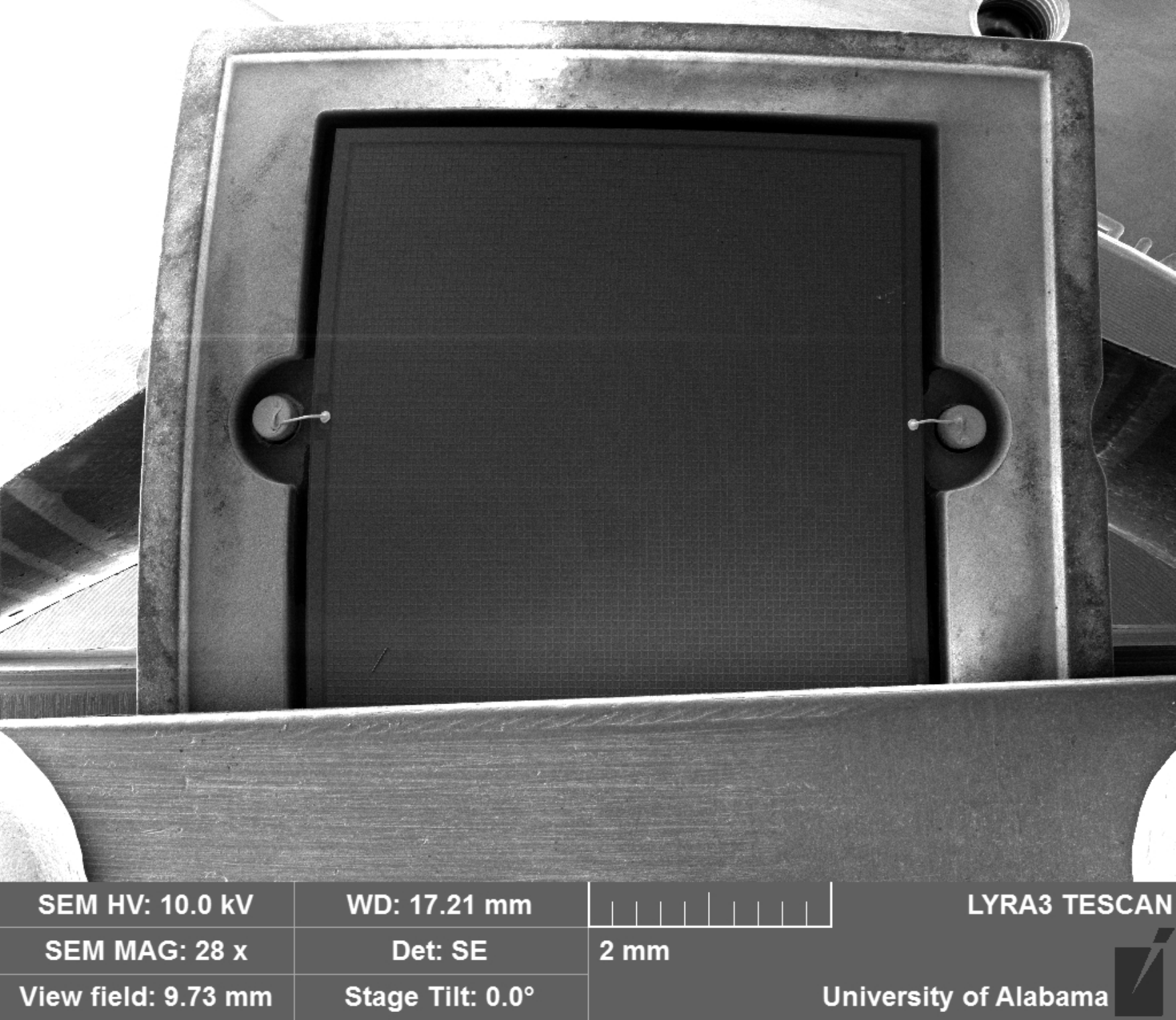}
  \caption{An SEM image of a Hamamatsu VUV4 SiPM (series: S13370).}
 \label{fig:vuv_whole}
\end{figure}

The nEXO collaboration studied the performance of both Hamamatsu and FBK SiPMs in external electric fields~\cite{nexo_paper1}, the absolute PDE of FBK and Hamamatsu SiPMs in vacuum and nitrogen gas~\cite{nexo_paper2, nexo_paper3}. 
More recently, the nEXO collaboration characterized the absolute PDE, rate of correlated avalanches, and dark count rate of the new generation of the Hamamatsu VUV4 SiPMs (series: S13370) in nitrogen gas~\cite{giacomo}. The SiPMs of this series are being considered by nEXO because they are directly sensitive to LXe scintillation light, which is measured to have a Gaussian emission spectrum with the center at 174.8 nm and the full width at half maximum of 10.2 nm~\cite{lxe_wave}. The SiPMs (Table~\ref{tab:sipms}) have a micro-cell pitch of 50 $\mu$m, an effective photosensitive area of 6x6 mm$^2$ with 60\% fill factor, and a typical breakdown voltage at room temperature of $\sim$52 V. The devices have a ceramic package and no window. 
\begin{table}[h]
\centering
\caption{Key parameters of the Hamamatsu VUV4 SiPMs (series: S13370) studied in this work.}
\label{tab:sipms}
\begin{tabular}{|l|c|}
\hline
Parameter & Value\\
\hline
Effective photosensitive area  & 6x6 mm$^2$ \\
Pixel pitch  & 50 $\mu$m \\
Fill factor  & 60\%  \\
Package  & Ceramic  \\
Window  & Unsealed \\
Breakdown voltage, 25\textdegree C  & $\sim$52 V\\
Breakdown voltage, -104\textdegree C  & $\sim$45 V\\
\hline
\end{tabular}
\end{table}
Two SiPMs of this series were measured to have PDE in the VUV of 14.8$\pm$2.8\% and 12.2$\pm$2.3\% at a mean wavelength of 189$\pm$7 nm, which is substantially lower than 24\% advertised by Hamamatsu~\cite{giacomo}. Hamamatsu SiPMs of a similar type, but consisting of four 6x6 mm$^2$ devices chained together (series: S13371), were also studied as possible candidates by DARWIN. Their rates of dark counts and cross-talk, and long term stability were characterized in LXe~\cite{baudis}. In this work we study the reflectivity and relative PDE of the VUV4 SiPMs (series: S13370) in LXe. Figure~\ref{fig:vuv_whole} shows an SEM image of a VUV4 SiPM.

The paper is organized as follows. First, the experimental setup and the measurement approach are described. Then, the systematic errors are discussed and estimated using Monte Carlo (MC) simulation and dedicated measurements. The validation of the approach is then described. Finally, the results of the VUV4 SiPMs measurements are presented, followed by concluding remarks. 

\section{Measurement approach}
\label{sec:meas}
Below we describe the features of the setup that are relevant for the current study.
\subsection{LIXO setup}
\begin{figure}[h]
 \centering
\includegraphics[width=0.45\textwidth]{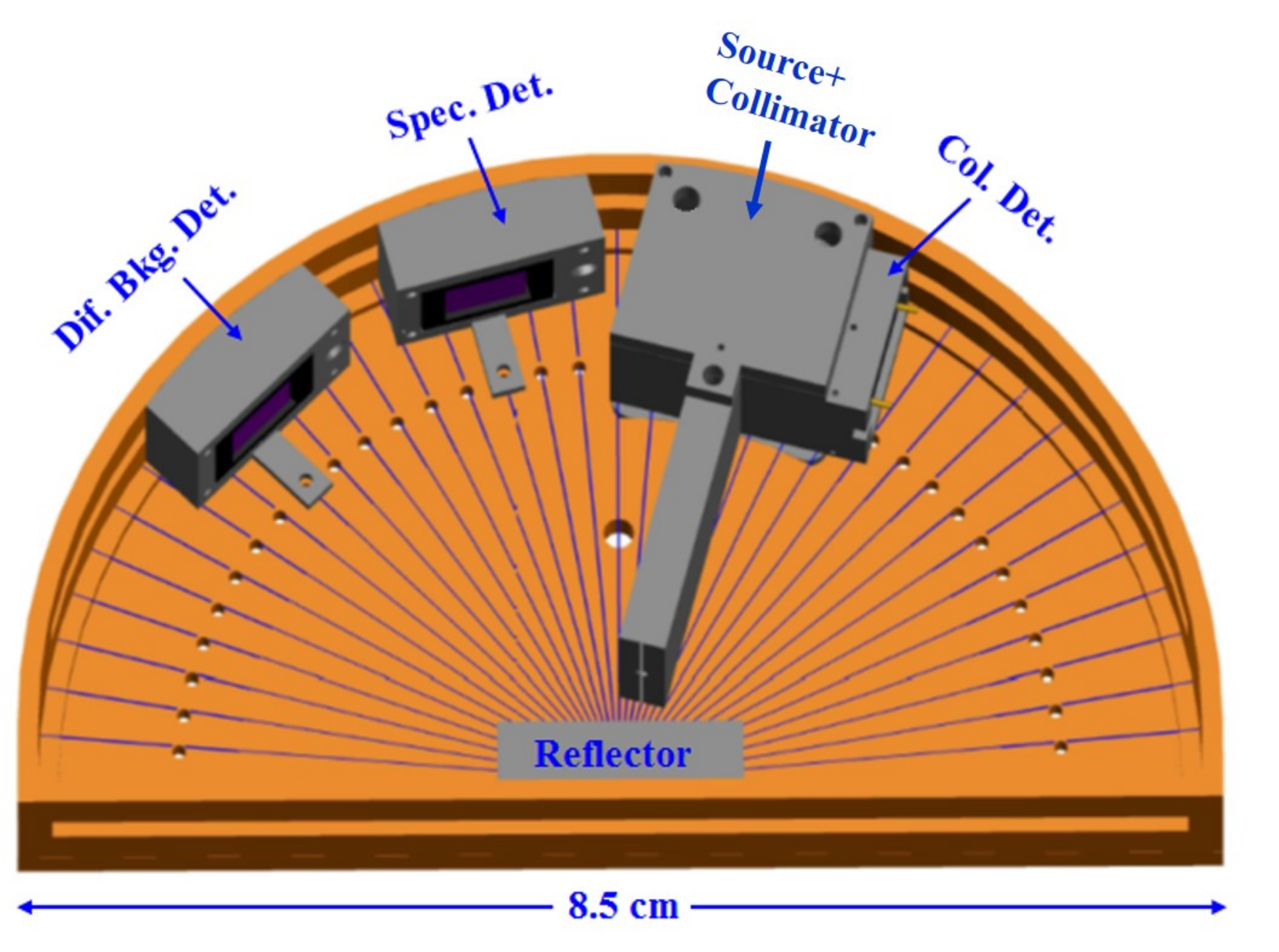}
\includegraphics[width=0.45\textwidth]{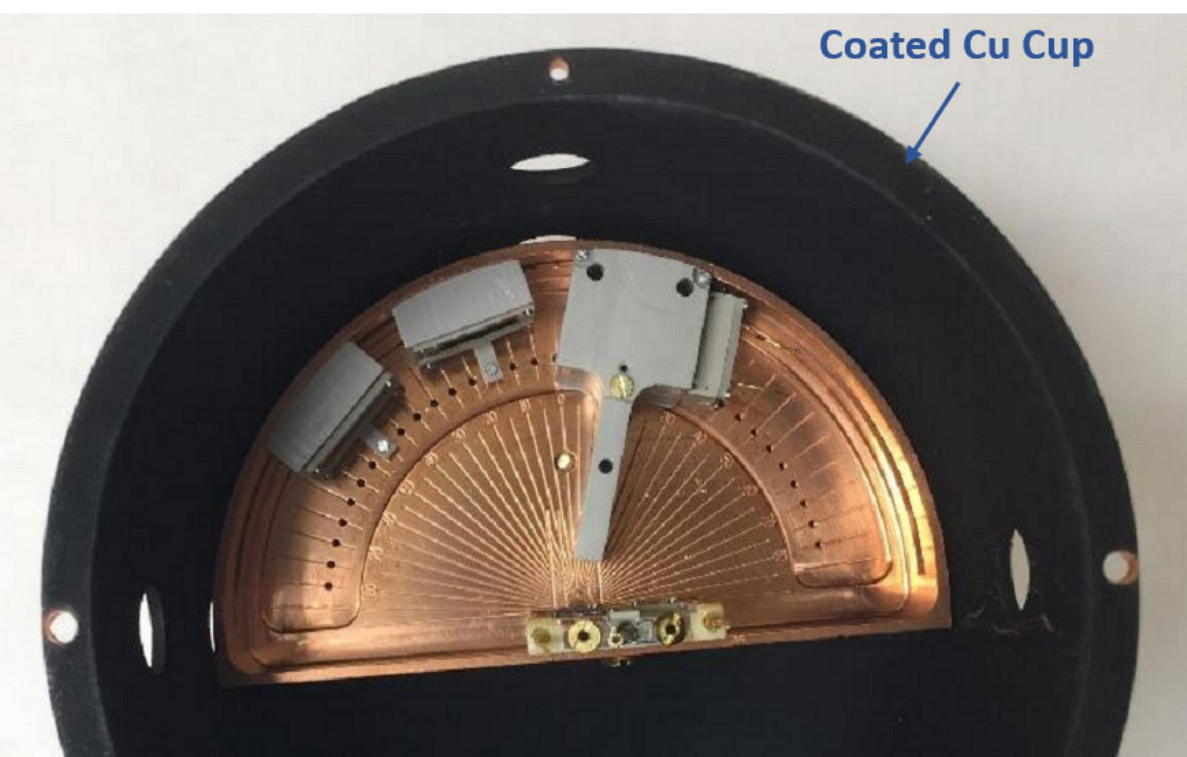}
  \caption{A CAD model (left) and a picture (right) of the LIXO setup. A semi-circular copper rail with angle degree marks is shown. A sample to be studied, which can be either a photosensor or a passive material, is marked as Reflector in the figure and is positioned such that its reflecting surface is at the center of the semi-circle. Two photodetectors are placed facing the reflector at 15 and 45 degrees with respect to the normal incidence on the reflector. The radioactive source is placed inside a collimator, which faces the sample from a fixed angular position of 15 degrees with respect to the normal incidence. Another photodetector is mounted on the side of the collimator, facing the radioactive source.}
 \label{fig:lixo}
\end{figure}

LIXO (LIquid Xenon Optics) is a setup built at the University of Alabama to measure the VUV reflectivity of different materials and PDE of different photosensors in LXe. It consists of a copper LXe chamber that can hold about 1~L of LXe during operation. The chamber is placed inside a vacuum insulated  cryostat. A semi-circular copper rail (Figure~\ref{fig:lixo}) with angle degree marks (every 5 degrees) is mounted inside the LXe chamber during measurements. The copper rail has an inner radius of 4.25 cm. A sample to be studied (either a photosensor or a passive material) is positioned such that its reflecting surface is at the center of the semi-circle (marked as Reflector in Figure~\ref{fig:lixo}). One or more photodetectors (in this work, Hamamatsu VUV4 SiPMs) are placed facing the reflector at fixed angular positions to detect scintillation light reflected from the sample. The holders for the detectors and the reflector are made from PEEK plastic. PEEK is chosen for its compatibility with LXe, low outgassing, and stability at low temperatures. Importantly, PEEK has been shown to strongly absorb UV radiation in the 300-400 nm wavelength range~\cite{peek}, and a direct measurement in the LIXO setup at LXe wavelengths (discussed later in Section~\ref{sec:MC}) indicates a small reflectivity of 2.6$\pm$0.6\%.

A $^{252}$Cf radioactive source is used to excite the LXe scintillation. 
$^{252}$Cf undergoes alpha and spontaneous fission (SF) decays. SF decays of $^{252}$Cf are less frequent than alpha decays (branching ratio of 3.1\%~\footnote{The apparent branching ratio may differ due to presence of other fissionable nuclides in the source, e.g., $^{248}$Cm}). However, SF produces substantially more scintillation light per decay, which could be useful when studying samples with low reflectivity, or when studying photosensors with low PDE. The radioactive source is placed inside a collimator, which faces the sample from a fixed angular position. The collimator is made of PEEK and has a bore diameter of 0.99 mm. 
 
Another photodetector is mounted on the side of the collimator, facing the radioactive source. The purpose of the collimator detector is two-fold. Firstly, it provides a trigger during reflectivity and PDE measurements. Secondly, it monitors the scintillation light level generated by the source and its stability. A Hamamatsu VUV4 SiPM is used as the collimator detector in this work.

The copper rail with the reflector, the detectors, and the source are attached to the bottom of a coated copper cup (the black cylinder on the right of Figure~\ref{fig:lixo}). 
The coated copper cup is fixed to the bottom of a vacuum flange that seals the top of the LXe chamber. The copper cup is coated with the MLS-85-SB black optical coating by AZ Technology, Inc.~\cite{az}. The coating has negligible reflectance at 175 nm. 

Once all components are assembled, the copper LXe chamber is closed, and the setup is evacuated to a residual pressure of $\sim$10$^{-6}$ Torr. The chamber is cooled and filled with LXe such that the reflector, the detectors, and the source are fully submerged. The xenon is purified on the way to the chamber using a 902 series MicroTorr SAES ambient temperature purifier (residual water contamination <0.1 ppb~\cite{saes}). 

Connection to the photosensors is provided by Kapton insulated cables~\cite{kapton}, which are routed outside the LXe chamber via an ultra-high vacuum (UHV) signal feedthrough. The signals are amplified and shaped by Cremat charge sensitive preamplifiers~\cite{cremat} and Cremat Gaussian shapers~\cite{cremat2}, respectively. The shaping time is 100 ns. The shaped signals are digitized by a 250 MS/s CAEN DT5725 digitizer~\cite{caen} and analyzed by a custom-made ROOT~\cite{root} based software. 

\subsection{Relative PDE measurement}
The absolute PDE of a SiPM is defined as the number of photoelectrons (p.e.) detected over the number of incident photons. Without knowing the absolute number of incident photons, it is still possible to compare the (relative) PDE of different SiPMs by comparing their corresponding number of p.e. detected when exposed to the same light source. A VUV4 SiPM to be studied is placed in the reflector position (Figure~\ref{fig:lixo}). A source is placed at a desired angle of incidence with respect to the reflector (with 0$^{\circ}$ corresponding to normal incidence). In later measurements, a second source, with its own collimator and a collimator detector, is added to measure at two incident angles during one liquefaction. 
In practice, the choice of the incident angle is currently limited to a range of 0 to 65 degrees for the PDE measurements and to a range of 15 to 65 degrees for the reflectivity measurements. The limitation is due to mechanical constraints at the low incident angles and a beam shadowing effect from the Hamamatsu ceramic SiPM holder at the large angles. The collimator design is chosen such that the incident and the specularly reflected light beams are fully contained by an active area of a 6x6 mm$^2$ detector, which is a common size of currently available SiPM detectors. 

\begin{figure}[h]
 \centering
\includegraphics[width=0.75\textwidth]{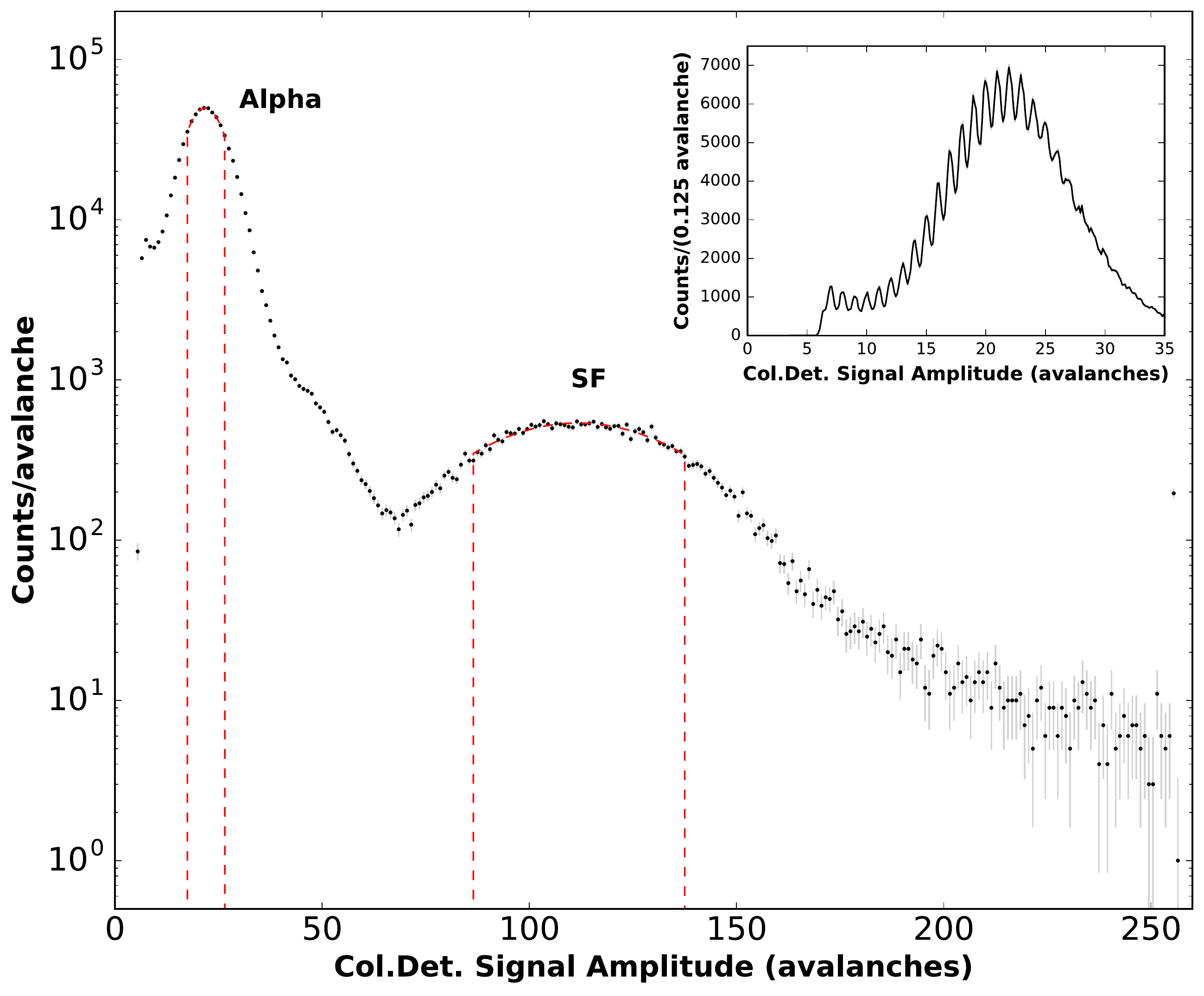}
  \caption{A spectrum of the xenon scintillation light excited by a $^{252}$Cf source, as seen by a collimator detector. Both the alpha and SF peaks are visible. The insert shows the alpha peak region in linear scale, with individual avalanches clearly resolved. The red dashed lines show the $\pm$1 $\sigma$ regions of the Gaussian fits to the alpha and SF peaks. Bias on the collimator detector is set to provide 1.3 V over-voltage.}
 \label{fig:spectrum}
\end{figure}

A measurement is performed by triggering on the collimator detector and recording signals from all detectors (the SiPM in the reflector position, one or both collimator detectors) for $\pm$2.5 $\si{\micro}$s around the trigger time. 
The trigger threshold is set just below the alpha peak in the collimator detector spectrum (Figure~\ref{fig:spectrum}). For the source used in the figure, the trigger threshold is $\sim$5 detected avalanches~\footnote{Detected avalanches, or avalanches for brevity, are used instead of p.e. to highlight the presence of cross-talk and afterpulsing in the spectrum.}
During the analysis stage, only those events on the detectors are considered that a) occur in the $\pm$0.1 $\si{\micro}$s coincidence window with the trigger and b) correspond to the alpha (or SF) signal in the triggering collimator detector. The alpha and SF signals are defined by fitting a Gaussian to the corresponding peak in the amplitude spectrum of the triggering collimator detector and placing a $\pm$1 Gaussian $\sigma$ cut around the peak position. When two sources are present, an additional cut is applied to exclude coincidences between signals from the triggering and non-triggering collimator detectors.

\begin{figure}[h!]
 \centering
\includegraphics[width=0.6\textwidth]{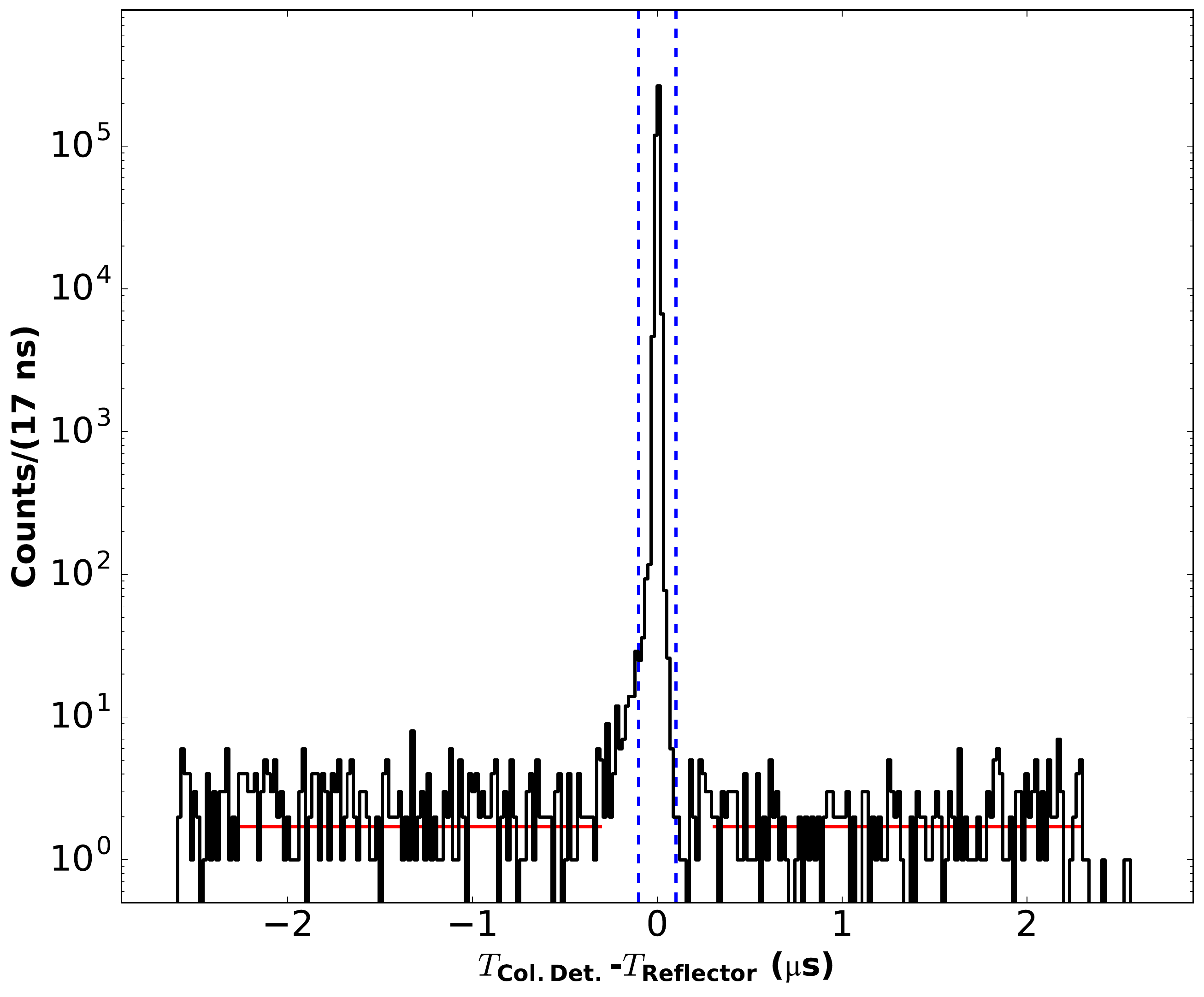}
  \caption{Distribution of event time differences between the SiPM in the reflector position and the alpha decays in the triggering collimator detector. The alpha decays are selected by applying a cut to the collimator detector's spectrum. The vertical dashed lines show the position of the coincidence cut used in the analysis. The horizontal lines show a constant fit to the uncorrelated coincidence rate. Bias on the reflector SiPM is set to provide 3.0 V over-voltage. Bias on the collimator detector is set to provide 1.3 V over-voltage.}
 \label{fig:dt}
\end{figure}

Figure~\ref{fig:dt} shows an example distribution of event time differences between the triggering collimator detector and the detector in the reflector position, $T_{\textup{Col.Det.}} - T_{\textup{Reflector}}$. A correlated coincidence peak is clearly seen. The sub-dominant flat background is due to random coincidences with the reflector SiPM's dark counts and, possibly, other uncorrelated light flashes in the LXe chamber (e.g., natural radioactivity, luminescence). 

\begin{figure}[h!]
 \centering
\includegraphics[width=0.60\textwidth]{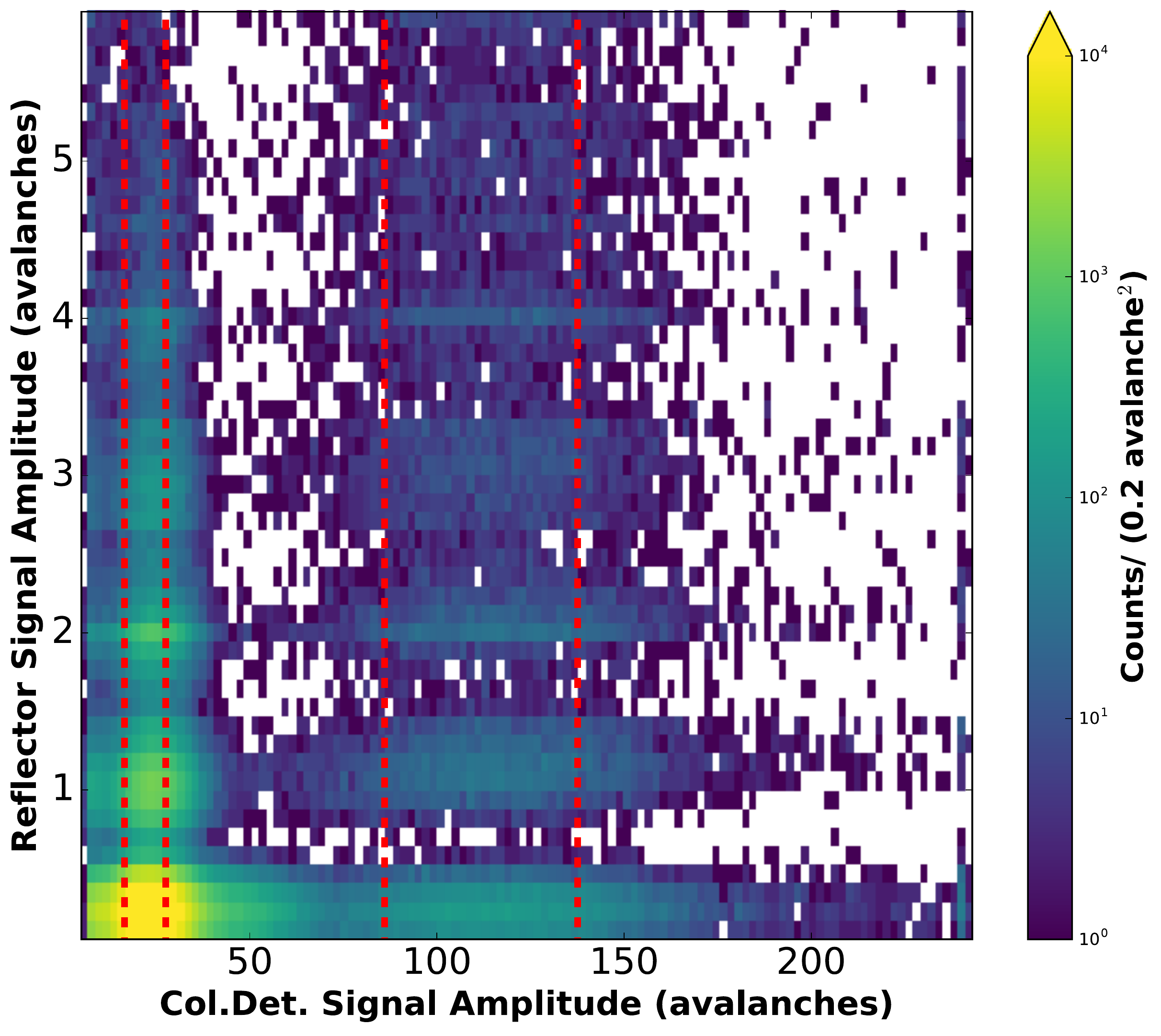}
 \caption{2D amplitude spectrum of coincidence events between the triggering collimator detector (X-axis) and the SiPM in the reflector position (Y-axis). The dashed vertical lines show positions of the cuts that select alpha and SF decays in the triggering collimator detector amplitude spectrum. Bias on the reflector SiPM is set to provide 3.0 V over-voltage. Bias on the collimator detector is set to provide 1.3 V over-voltage.}
 \label{fig:2d}
\end{figure}

An example 2D amplitude distribution of events in the coincidence peak is shown in Figure~\ref{fig:2d}. The X-axis corresponds to the signal amplitude of the triggering collimator detector. The Y-axis corresponds to the signal amplitude of the SiPM detector in the reflector position. The dashed vertical lines show positions of the cuts that select alpha and SF events in the triggering collimator detector amplitude spectrum.

\begin{figure}[h!]
 \centering
\includegraphics[width=0.6\textwidth]{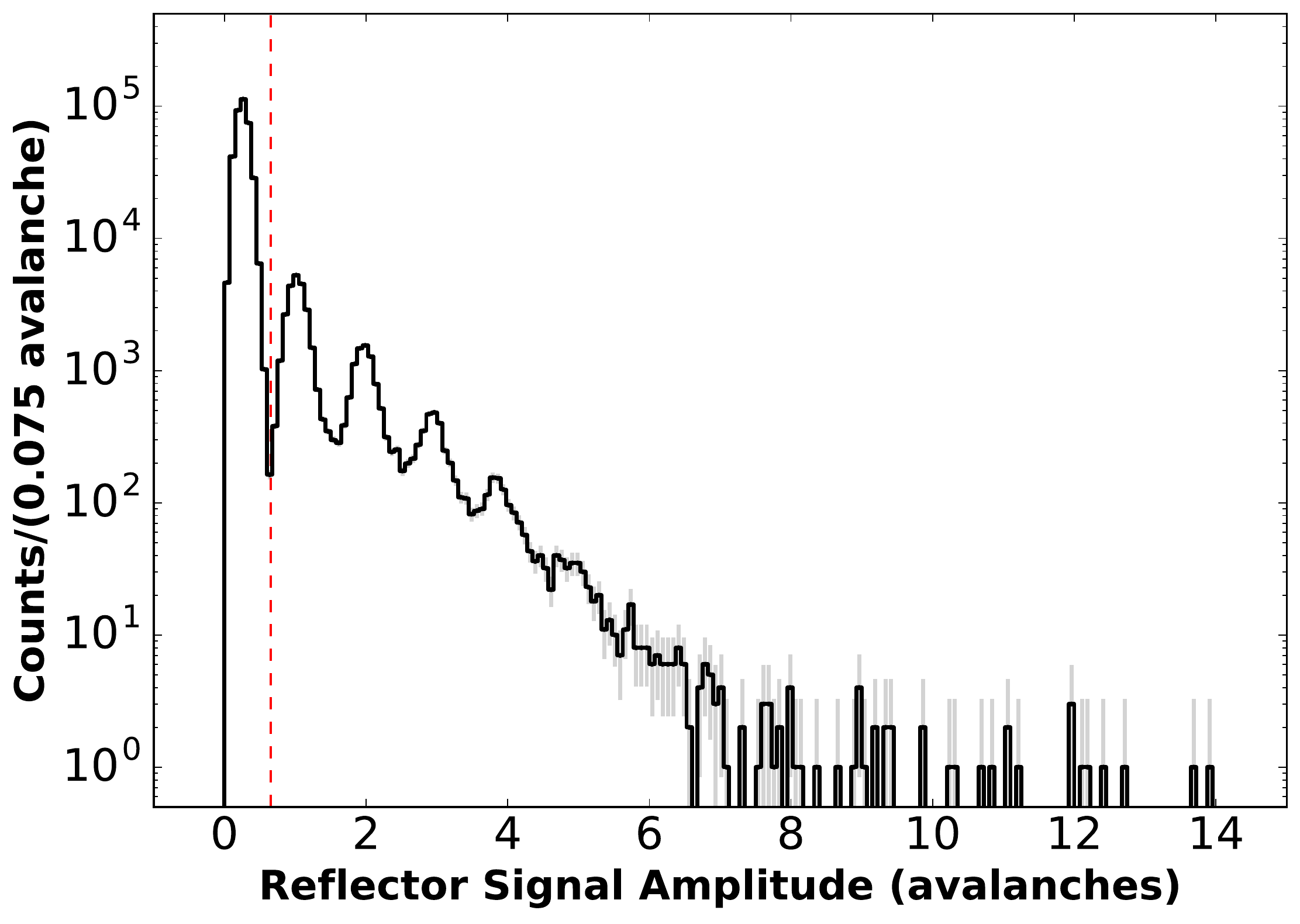}
  \caption{Amplitude distribution of events in the VUV4 SiPM placed in the reflector position. Only those events that coincide with the alpha decays in the triggering collimator detector are selected. The red dashed vertical line shows the position of a cut that separates the events with zero avalanches. Statistical errors are shown in gray. Bias on the SiPM is set to provide 3.0 V over-voltage.}
 \label{fig:spec}
\end{figure}

An example amplitude distribution of events in the SiPM in the reflector position that coincide with the alpha decays in the triggering collimator detector is shown in Figure~\ref{fig:spec}. Effectively, this distribution is a projection on the Y-axis of the 2D distribution shown in Figure~\ref{fig:2d} after applying the alpha decay cut.

As can be seen from Figure~\ref{fig:spec}, the collimation reduces the amount of light at the reflector down to a few photons. Assuming Poisson statistics, the average number of p.e. seen by a photodetector in the reflector position, $\mu^\textup{R}$, is then related to the fraction of the selected triggers in which no pulses have been detected by that detector:
\begin{equation}
\mu^\textup{R} = -\log(N_0/N),
\label{eq:mu}
\end{equation}
where $N_0$ is the number of events with zero avalanches (the integral of the first peak in Figure~\ref{fig:spec}) and $N$ is the total number of selected events.

The $\mu^\textup{R}$ determination method is not affected by cross-talk and afterpulsing of SiPMs~\cite{giacomo,meg}. The method is in principle affected by dark counts in the SiPM and other sources of random coincidences. However, as can be seen from Figure~\ref{fig:dt}, their contribution to the coincidence window is negligible ($<$0.1\%) in practice. More information about methods of measuring PDE of SiPMs can be found Refs.~\cite{pde1,pde2,pde3,pde4,pde5}.

A yet to be identified source of light apparently produces correlated coincidences in all submerged detectors. This leads to an increase of the observed average number of p.e.s of all detectors. The amount of the increase, which we denote as $\mu^{\textup{D}}$, does not vary significantly from one liquefaction to another. It is not found to have a significant dependence on the angle of incidence or the type of a sample placed as a reflector, suggesting a nearly isotropic (diffuse) nature. Some of the possible sources of the diffuse background that are currently being investigated are cosmic muons, unexpected parasitic reflections, and neutrons. 

The $\mu^\textup{R}$ value can be measured as described above for different photosensors at different over-voltages and angles of incidence. As long as the light level produced by the source is stable, the PDE of different photosensors can be compared using their relative values of the $\mu^\textup{R}$. In other words, for two photodetectors, denoted by indices ``1'' and ``2'', at the same over-voltage, and assuming a constant level of the light source, the ratio of the detectors' PDE is given by:
\begin{equation}
\frac{PDE_1}{PDE_2} = \frac{\mu^\textup{R}_1-\mu^\textup{D}}{\mu^\textup{R}_2-\mu^\textup{D}}
\label{eq:pde}
\end{equation}
When comparing measurements obtained with different sources, an additional constant scaling factor is introduced. The scaling factor is independent of the angle of incidence and accounts for differences between the two used $^{252}$Cf sources. For the two sources used in this work, the scaling factor is measured to be 1.409$\pm$0.027. 
While it is possible to estimate the absolute PDE with additional effort, this is outside the scope of this work. 

\subsection{Reflectivity measurement}
A sample to be measured is placed in the reflector position (Figure~\ref{fig:lixo}). A source is placed at a chosen angle of incidence with respect to the reflector. A detector  -- a ``specular detector'' -- is placed at the same (reflected) angle as the source to measure specular reflectivity. One or more additional detectors -- ``diffuse background detectors'' -- are placed at angles different from the specular one. VUV4 SiPMs are used as specular and diffuse background detectors in this work.
The diffuse background detectors measure photons that are produced by the correlated coincidence background. They also register photons that are diffusely reflected from the sample. However, given the solid angle and the observed level of the background, the diffuse reflection contribution is not discernible for samples with low diffuse reflectivity, such as SiPMs. 
Importantly, both the specular and diffuse detectors have their $\mu^\textup{R}$ measured at normal incidence in advance of the reflectivity measurement. 

During a reflectivity measurement, signals from the specular and diffuse background detectors are recorded when a trigger is issued. Their mean numbers of p.e.  --- $\mu^\textup{S}$ ($\mu^\textup{D}$) for specular (diffuse) detectors --- are determined analogously to $\mu^\textup{R}$ (Eq.~\eqref{eq:mu}). 

\begin{figure}[h]
 \centering
\includegraphics[width=0.6\textwidth]{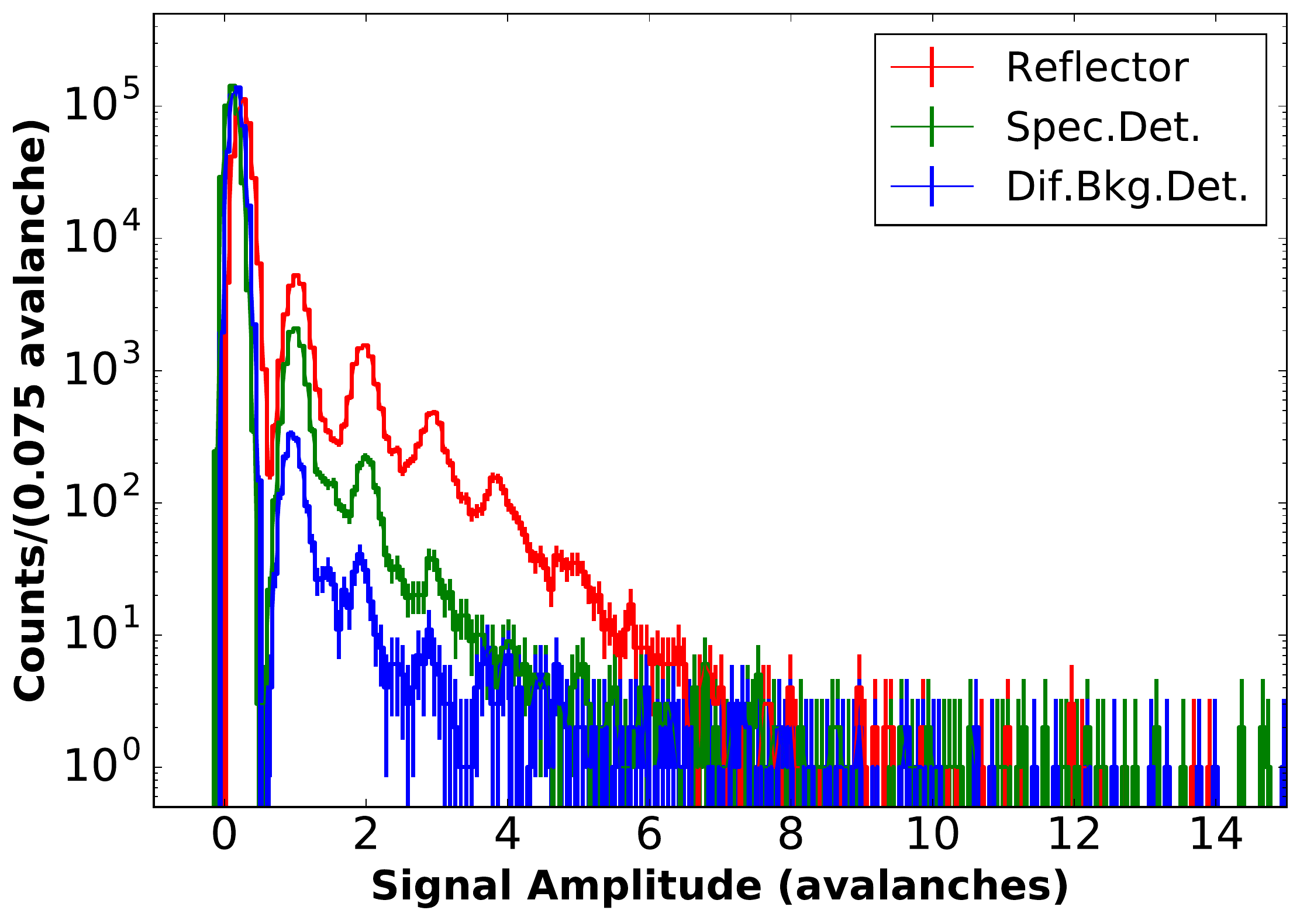}
  \caption{Amplitude distribution of events in the detectors placed in the reflector (red), the specular detector (green), and the diffuse background detector (blue) positions. Only those events that coincide with the alpha decays in the triggering collimator detector are selected. Bias on each of the detectors is set to provide 3.0 V over-voltage.}
 \label{fig:1d}
\end{figure}

Example amplitude distributions for a single reflectivity measurement is shown in Figure~\ref{fig:1d}. In this example, the source was positioned (see Figure~\ref{fig:lixo}) at 15$^{\circ}$ angle of incidence on a VUV4 SiPM that was in the reflector position. The specular detector was at 15$^{\circ}$ reflected angle, and the diffuse background detector was at 45$^{\circ}$ reflected angle. As can be seen from Figure~\ref{fig:1d}, the average number of p.e. is the largest on the reflector and the smallest on the diffuse background detector, as expected. 

If one assigns an index ``1'' to the sample in the reflector position, an index ``2'' to the specular detector, then the specular reflectivity of the sample, $R_1$, can be found as follows:
\begin{equation}
R_1 =\frac{\mu^\textup{S}_2-\mu^\textup{D}}{\mu^\textup{R(0)}_2-\mu^\textup{D}},
\label{eq:r}
\end{equation}
where $\mu^\textup{R(0)}_2$ refers to the mean number of p.e. on the specular detector measured (prior to the reflectivity measurement) in the reflector position with the source at normal incidence. 
Eq.~\eqref{eq:r} can be used to study specular reflectivity in LXe of both photosensors and passive materials. 
It assumes that the light emitted by the source is stable during different liquefactions, that the light specularly reflected by a sample is fully contained on a single 6x6 mm$^2$ detector, and that other systematic effects (parasitic reflections, Rayleigh scattering) are negligible. These assumptions are studied, and the related systematic errors are quantified, in the following section.

\section{Systematic effects}

\subsection{Optical simulation}
\label{sec:MC}
The LIXO setup is modeled in Chroma, an ultra-fast photon MC software that runs on graphics processing units (GPUs)~\cite{chroma}. Chroma was chosen for this work mainly because it allows one to input a fully-detailed geometry of a setup directly from a CAD file. It models photon generation, propagation, Rayleigh scattering, specular and diffuse reflection, absorption, and detection. 

MC studies were conducted to investigate the following effects:
\begin{enumerate}
\item Beam collimation

For this study only the geometric divergence of the beam is considered, so the effects of absorption, Rayleigh scattering, and parasitic reflections are turned off. 
Several millions of photons per decay are generated at the source position inside the collimator. The photons' initial direction distribution is isotropic. Photons that escape the collimator's 0.99 mm diameter bore and hit the reflector are counted as $N_{\textup{in}}$. In a separate simulation with the same number of initial photons, all photons that hit the reflector get specularly reflected. The number of the reflected photons that hit the 6x6 mm$^2$ specular detector are counted as $N_{\textup{det}}$. We then define ``containment'', $C$, as the ratio of $N_{\textup{det}}$ to $N_{\textup{in}}$:
\begin{equation}
C = \frac{N_{\textup{det}}}{N_{\textup{in}}}
\label{eq:c}
\end{equation}
Figure~\ref{fig:c} shows positions of simulated photon hits on the reflector (left) and the specular detector (right) for the case of 15$^{\circ}$ incident angle. 
\begin{figure}[h]
 \centering
 \includegraphics[width=0.85\textwidth]{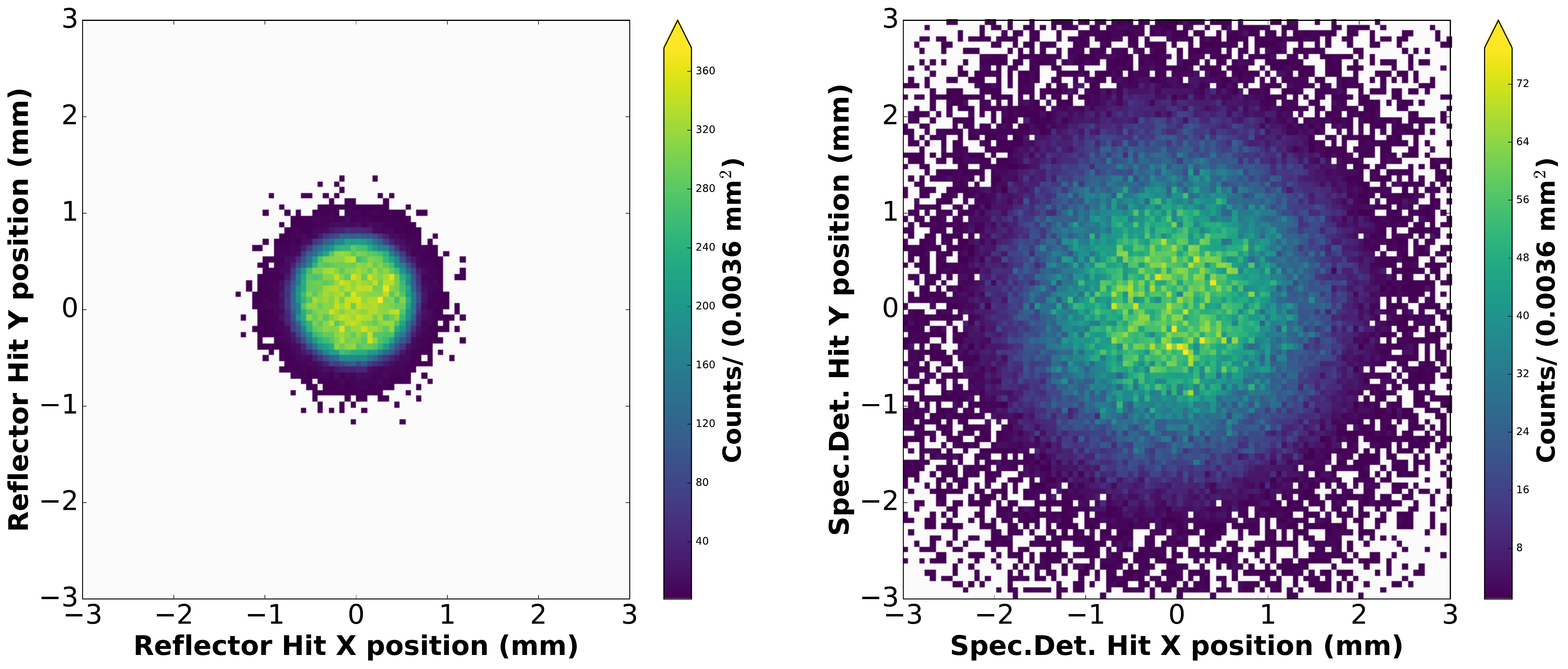}
  \caption{Distribution of simulated photon hits on the reflector (left) and the specular detector (right). The source is positioned at 15$^{\circ}$ incidence on the reflector. In both cases the light beam is fully contained in a 6x6 mm$^2$ area. Only the geometric divergence of the beam is considered, so the effects of absorption, Rayleigh scattering, and parasitic reflections are turned off.}
 \label{fig:c}
\end{figure}
The diameter of the light beam increases by roughly a factor of two on the way from the collimator to the reflector, and then by another factor of two on the way to the specular detector. Nevertheless, the reflected light beam is effectively contained on the specular detector. 

Apart from the diameter and the length of the bore, the containment is a function of the reflectivity of the collimator's material.  
A sample from the particular batch of PEEK that was used to manufacture the LIXO's collimators has been measured in vacuum at room temperature and found to have $\lesssim$1\% reflectivity at 175 nm~\cite{fatemeh}. The same batch of PEEK was also measured in LIXO and found to have 2.6$\pm$0.6\% reflectivity in LXe. The containment value averaged between the two PEEK measurements is 99.3$\pm$0.4(stat)$\pm$0.7(sys)\%. The systematic error is the spread between the two measurements.

\item Rayleigh scattering

The Rayleigh scattering systematically reduces the apparent reflectivity of a sample. We repeat the previous study with the Rayleigh scattering effect now turned on and evaluate the containment as a function of the scattering length. Figure~\ref{fig:ray} shows the results of the study. 
\begin{figure}[h]
 \centering
\includegraphics[width=0.7\textwidth]{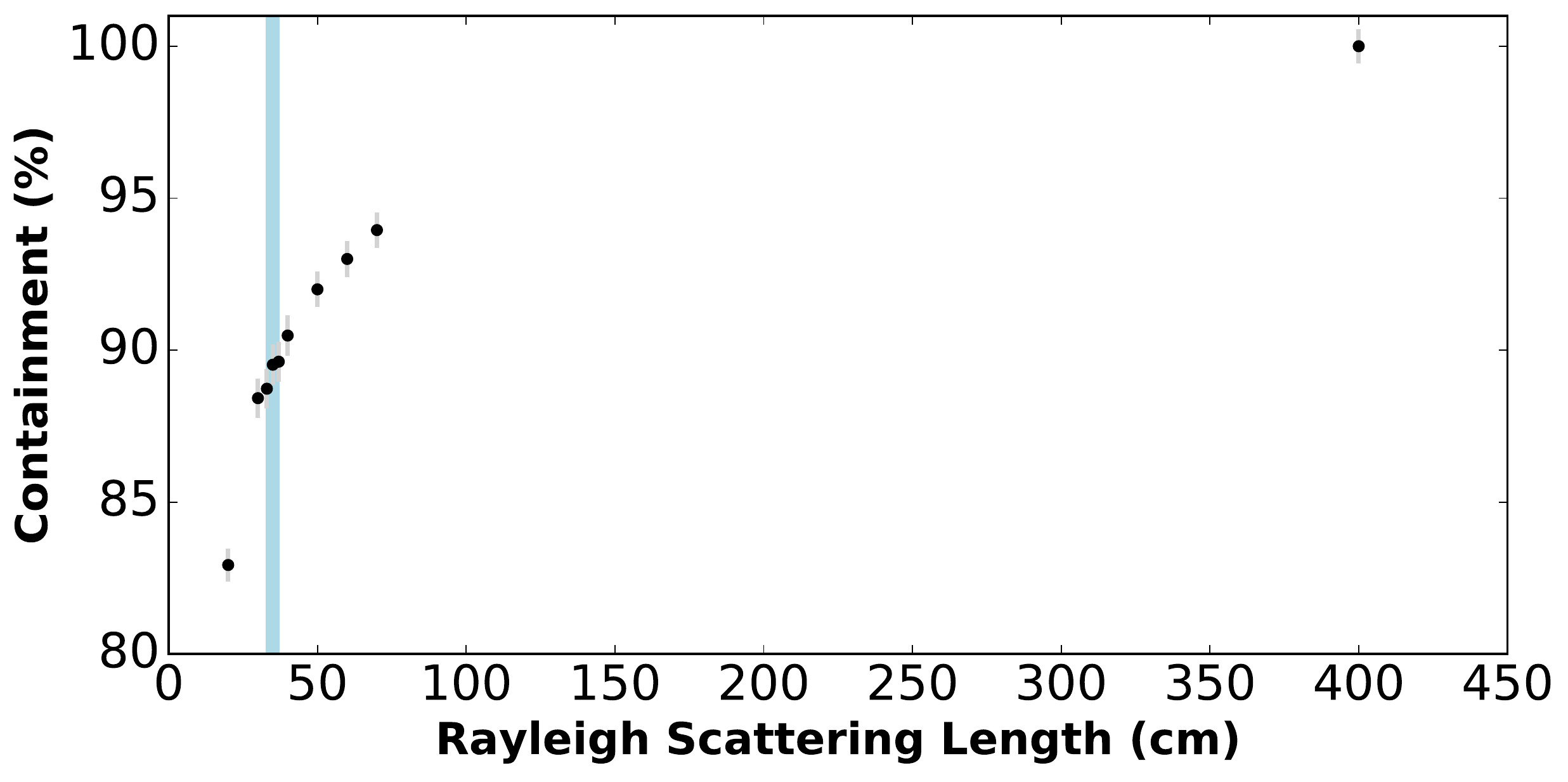}
  \caption{The containment of the specularly reflected beam on a 6x6 mm$^2$ specular detector as a function of the Rayleigh scattering length. The vertical band corresponds to the expected value of the scattering length~\cite{rayleigh}.}
 \label{fig:ray}
\end{figure}
As the scattering length decreases, the containment also decreases, as expected. The error bars in the figure are statistical. The vertical band corresponds to the expected value of the Rayleigh scattering length in LXe, 35$\pm$2 cm~\cite{rayleigh}. The corresponding containment value is 90.1$\pm$0.7(stat)$\pm$1.8(sys)\%. 

\item Photon absorption

Xenon scintillation photons propagating in LXe can get absorbed by impurities, most notably water~\cite{absorption}. This effect systematically reduces the apparent reflectivity of a sample. 

Measuring the absorption length in LXe requires large detectors and an ability to disentangle the effects of absorption and scattering. Baldini et al.~\cite{absorption} have measured the absorption length using a 100 liter LXe detector. When acrylic, which is known to absorb and release water, was used in the setup, the initial absorption length before the purification was found to be $\sim$12 cm. After the acrylic parts were replaced with a low outgassing material, the absorption length before purification was found to be $\sim$30 cm. In both cases, the absorption length increased to $\sim$100 cm after the purification with the PS-15 SAES getter~\cite{saes} and the Oxisorb molecular filter~\cite{oxisorb}.

The LIXO's gas handling system is constructed using all-metal components and electro-polished stainless steel tubing. 
Only cleaned, baked LXe compatible materials are used inside the LXe chamber (PEEK, Kapton, copper, stainless steel). 
The LXe chamber and the gas handling system are evacuated to pressure on the order of 10$^{-6}$ Torr or better before each liquefaction. Only oil-free vacuum pumps are used in the lab. A stock of xenon with the 99.999\% purity Research Grade was loaded in the system. A 902 series MicroTorr~\cite{saes} SAES ambient temperature purifier is used to purify the xenon before each liquefaction. The MicroTorr purifier removes water at least as efficiently as the PS-15 getter (the limit specified on the SAES website for our model is $<$0.1 ppb, while for the PS series a $<$1 ppb limit is shown). 

Based on the above, we assume that the absorption length in LIXO is close to the purified value ($\sim$100 cm) reported by Baldini et al. We further assume a $\pm$40 cm uncertainty on the absorption length. The corresponding containment value is 96.6$\pm$0.6(stat)$\pm$2.7(sys)\%. 
Figure~\ref{fig:abs} shows the containment as a function of the absorption length and can be used to estimate the containment for other assumed values of the absorption length. While we are not able to measure the absorption length in LIXO \begin{it}in situ\end{it}, we can validate our assumptions by measuring samples with known expected reflectivity. 
\begin{figure}[h]
 \centering
\includegraphics[width=0.7\textwidth]{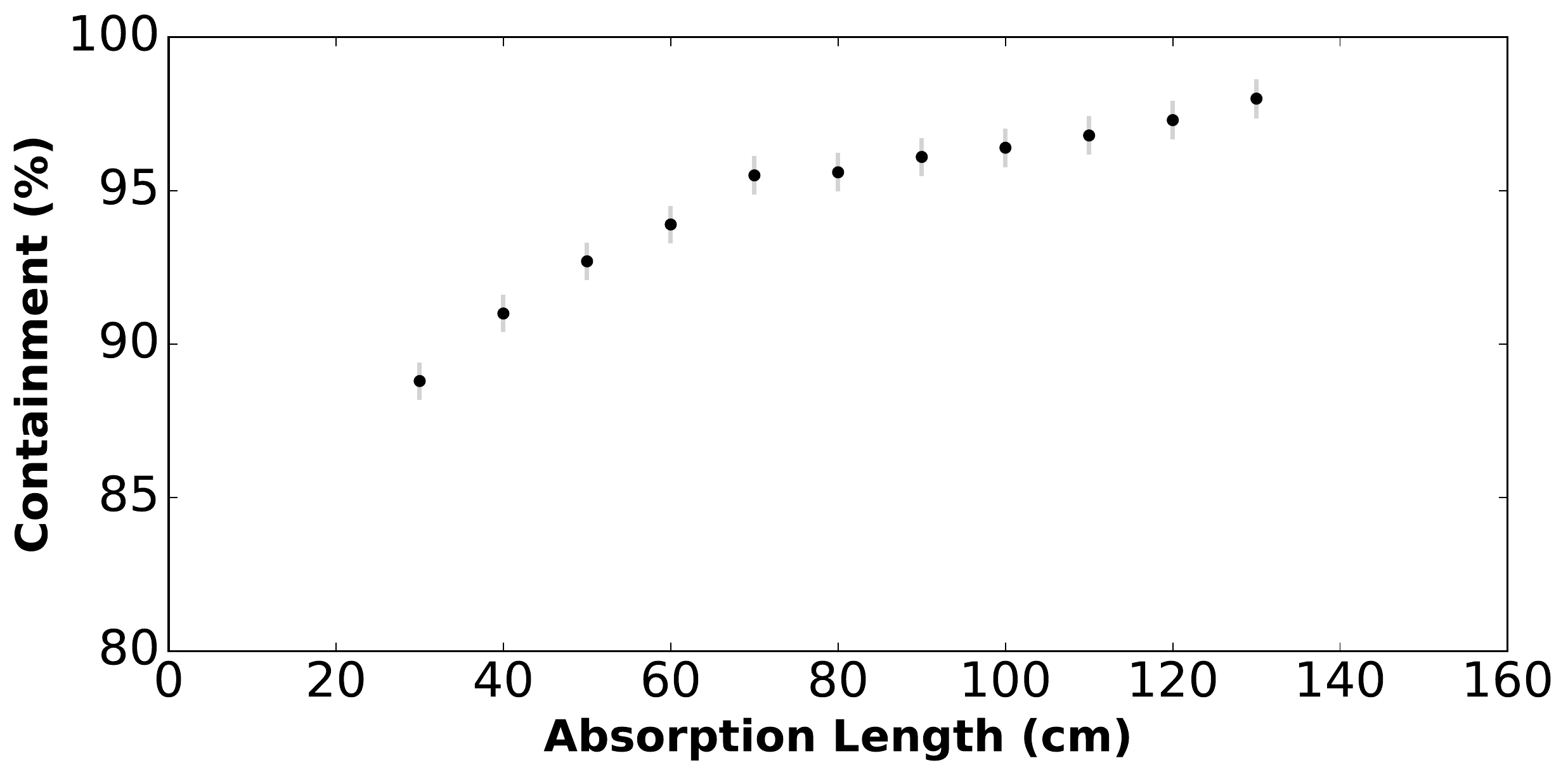}
  \caption{The containment of the specularly reflected beam on a 6x6 mm$^2$ specular detector as a function of the absorption length.}
 \label{fig:abs}
\end{figure}

\item Parasitic reflections

The contribution of parasitic, or spurious, reflections in LIXO is minimized by design. Materials that are known to be good reflectors in the VUV (Teflon~\cite{Neves_2017}, aluminum~\cite{al_r}) are avoided inside the LXe chamber, while materials that are known to be good absorbers (PEEK~\cite{peek}) are used commonly. An exception to this rule is copper, which can have a noticeable reflectivity in the VUV. Copper is used as a material for the LXe chamber to ensure good temperature uniformity. At the same time, the large solid angle of the LXe chamber could be a significant contributor to the parasitic reflections. To exclude this, the LIXO rail assembly is enclosed in a copper cup that is coated with a dedicated VUV absorbing coating (see picture on the right of Figure~\ref{fig:lixo}). We consider the semi-circular (uncoated) copper rail as a remaining potential contributor to the parasitic reflections. 

The reflectivity of copper at VUV wavelengths is rather uncertain and depends on the surface conditions that may change with time, e.g., due to oxidation~\cite{silva2007}. So we repeat the previous study with the Rayleigh scattering length now set to the expected value. We then turn on the copper reflectivity and scan a wide range of specular and diffuse reflectivity values. The containment, which in this case normalized to 100\% for non-reflective copper, is plotted as a function of specular and diffuse reflectivity of the rail in Figure~\ref{fig:cu}. 
\begin{figure}[h]
 \centering
 \includegraphics[width=0.95\textwidth]{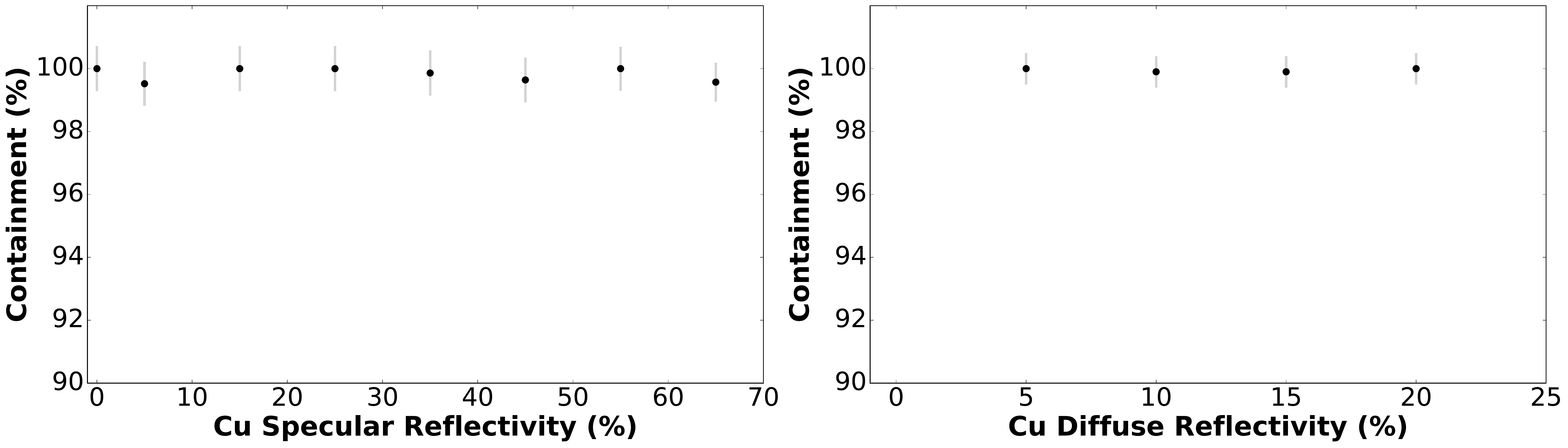}
  \caption{The containment of the specularly reflected beam on a 6x6 mm$^2$ detector as a function of the copper rail's reflectivity.}
 \label{fig:cu}
\end{figure}
The containment is found to be practically unaffected by reflections off the copper for the reflectivity values considered. We conclude that the effect of parasitic reflections in LIXO is sub-dominant. 

\item Overall systematic effect on the reflectivity measurement due to all modeled effects

For this study, a LIXO reflectivity measurement is modeled with all the considered effects turned on and set to their expected values. A VUV4 SiPM is modeled in the reflector position. Another VUV4 SiPM is modeled in the specular detector position. The SiPMs' reflectivity is set to 30\% specular and 15\% diffuse. The Rayleigh scattering length is set to 35 cm and the absorption length is set 100 cm. The reflectivity of copper is set to 35\% specular~\cite{cu_refl2009} and 10\% diffuse (based on an educated guess). The apparent specular reflectivity of the modeled sample is found to be 25.8\%. In the absence of any systematic effects, the apparent reflectivity should be equal to the true modeled value. The overall effect is, thus, 4.2 abs.\% (absolute), or 16.2 rel.\% (relative). The effect is dominated by the Rayleigh scattering in LXe. The relative effect does not depend on the absolute value of the modeled reflectivity, or the incident angle. Adding (in quadrature) the assumed uncertainties of the beam collimation, the Rayleigh scattering length, and the absorption length, the total systematic uncertainty is $\pm$3.2 rel.\%. 
\end{enumerate}

\subsection{Summary of systematic errors and corrections}
\label{sec:sys_summary}
Below is the list of systematic effects considered in this work:
\begin{itemize}

\item Effects modeled in the MC (Rayleigh scatter, adsorption, parasitic reflections, imperfect collimation).

The studies described in the previous sub-section suggest that a measurement of any sample obtained using Eq.~\eqref{eq:r} would result in a systematically lower value of reflectivity. This effect does not affect relative reflectivity values of different samples, but it lowers the absolute scale by 16.2$\pm$3.2 rel.\%. We therefore correct all reflectivity measurements by this amount. 

\item Light level stability.


We track the stability of the light level by monitoring the mean positions of the alpha and SF peaks in the collimator detector spectra collected in different measurements. Different measurements correspond to different xenon liquefaction-recovery cycles, separated by days or weeks. The LXe chamber is opened (e.g., to change the sample) before each measurement. Figure~\ref{fig:light} shows the light stability during 11 measurement cycles used in this work. 
\begin{figure}[h]
 \centering
\includegraphics[width=0.95\textwidth]{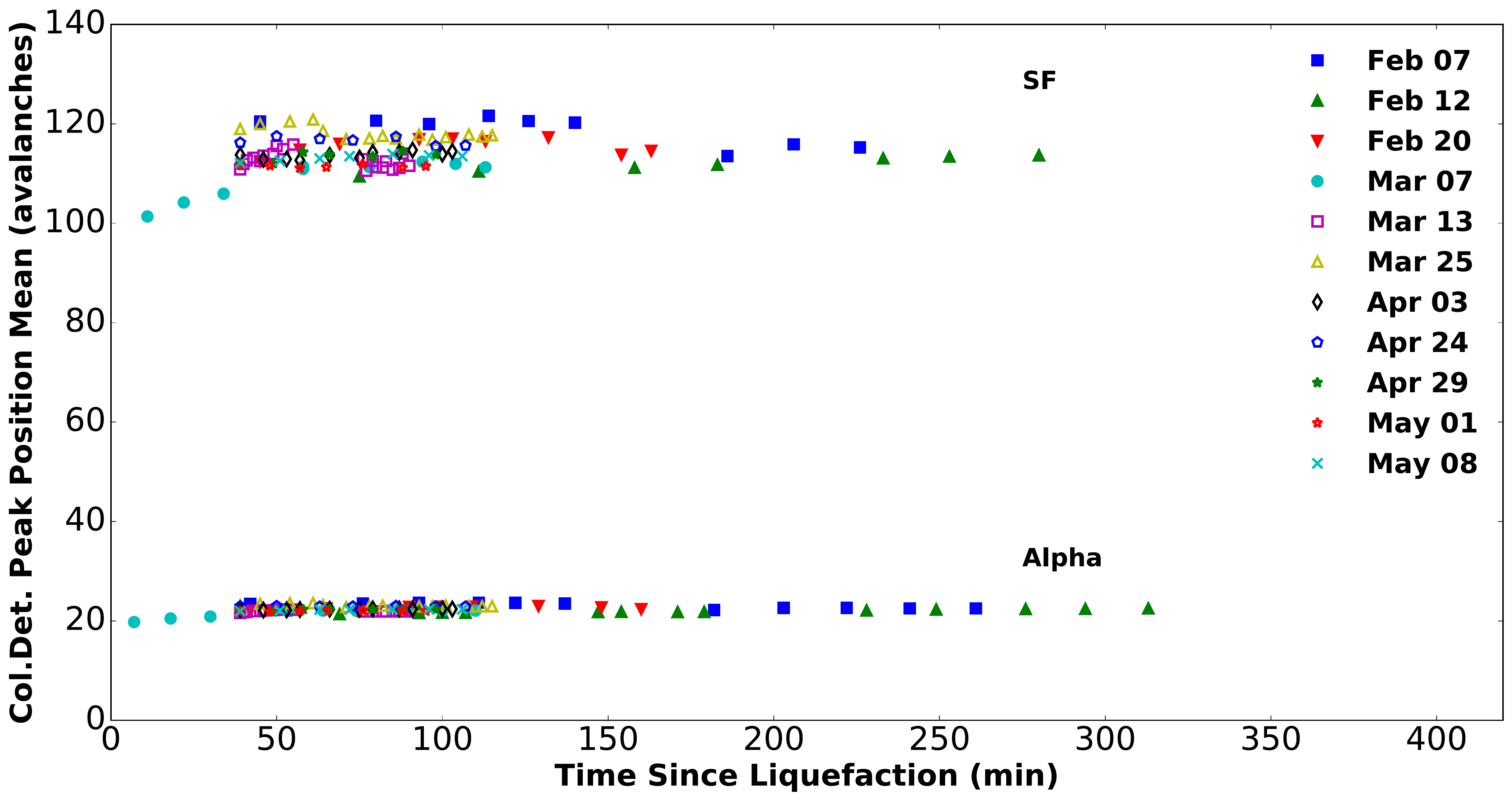}
  \caption{Mean positions of the alpha and SF peaks in the collimator detector spectra. X-axis shows time since liquefaction. Y-axis shows the mean peak position, as determined by a Gaussian fit to the corresponding peak in the amplitude distribution.}
 \label{fig:light}
\end{figure}
A measurement cycle typically lasts 2-4 hours. During the first half an hour after the liquefaction, the temperature and pressure in the LXe chamber stabilize, while the light level may gradually change by 10-15\% (see the March 7 data in Figure~\ref{fig:light}). We do not take reflectivity or PDE data during the first half an hour. 
The 11 measurement cycles shown in Figure~\ref{fig:light} were conducted over a period of three months. For these 11 measurement cycles the standard deviation (Std) from the average of the alpha (SF) peak position is found to be $\pm$2.3\% ($\pm$2.8\%).

To check if the light stability observed by a collimator detector correlates with the light stability in the reflector position we repeatedly measure the $\mu^\textup{R}$ (Eq.~\eqref{eq:mu}) for one of the three VUV4 SiPMs (serial number 141) studied in this work. Table~\ref{table:mu_a} summarizes the results of this check.
Each of the four measurements is performed at normal incidence. The alpha cut is used in the analysis. The SiPM was operated at 3 V over-voltage. Each of the four data points are collected in a separate liquefaction cycle, separated by several days to several weeks. The numbers in the round brackets show the statistical uncertainty in the last digit.
\begin{table}[h]
\centering
\caption{Stability check results. Mean number of p.e. on on a SiPM placed in the reflector position is measured repeatedly. Each of the four measurements is performed at normal incidence. The alpha cut is used in the analysis. The SiPM was operated at 3 V over-voltage. Each of the four data points are collected in a separate liquefaction cycle, separated by several days to several weeks. The numbers in the round brackets show the statistical uncertainty in the last digit.}
\label{table:mu_a}
\begin{tabular}{|l|c|}
\hline
Measurement\# & $\mu^\textup{R}_{141}$, p.e.\\
\hline
1  & 0.0938(6) \\
2  & 0.0936(6) \\
3  & 0.0958(7)  \\
4  & 0.0939(7)  \\
\hline
\hline
Mean$\pm$Std: &  0.0943$\pm$0.0009 \\
\textbf{Std/Mean}: & $\pm$\textbf{0.9}\% \\
\hline
\end{tabular}
\end{table}
We therefore assume a slightly conservative $\pm$3\% contribution to the uncertainty of each measurement of $\mu$ due to the light level stability.

\item Coincidence background.

We measure this contribution repeatedly by placing detectors at various angular positions away from the specular one. We repeat such measurements for different samples in the reflector position -- different SiPMs detectors, a silicon wafer (expected specular reflectance of $\sim$50\%), and a PEEK sample (expected specular reflectivity of close to 0\%). Table~\ref{table:difbkg} summarizes the results, suggesting a relatively stable value of the background from one liquefaction to another and an absence of a strong dependence on the angle of incidence or the type of a sample placed as a reflector. The diffuse background constitutes ca. 3\% of the incident flux for the alpha cut, but as much as 50\% for the S.F. cut, making the latter analysis much less accurate. Until the diffuse background is understood and eliminated, the alpha cut analysis is used for all measurements. 
\begin{table}[h]
\centering
\caption{Diffuse background measurements. Correlated coincidence background is measured by placing detectors at various non-specular angular positions. Results with different samples in the reflector position are shown. No strong dependence on the angle or sample type is observed. An average value of the coincidence background is also shown.}
\label{table:difbkg}
\begin{tabular}{|c|c|c|c|}
\hline
Non-specular angle, degree & Reflecting sample & $\mu^\textup{D}$, p.e. (alpha cut) & $\mu^\textup{D}$, p.e. (SF cut)\\
\hline
\hline
15 & VUV4, 141 & 0.0042(1) & 0.506(7) \\
45 & VUV4, 141 & 0.0040(1) & 0.487(6) \\
60 & VUV4, 140 & 0.0037(1) & 0.483(9) \\
45 & VUV4, 139 & 0.0047(1) & 0.499(5)\\
10 & Si wafer & 0.0029(1) & 0.435(6) \\
15 & Si wafer & 0.0029(1) & 0.432(6) \\
35 & Si wafer & 0.0032(1) & 0.465(6) \\
40 & Si wafer & 0.0033(1) & 0.512(7) \\
40 & Si wafer & 0.0034(1) & 0.477(7) \\
65 & Si wafer & 0.0036(1) & 0.507(7) \\
40 & PEEK & 0.0039(1) & 0.402(11) \\
65 & PEEK & 0.0033(1) & 0.348(10) \\
\hline
\hline
\multicolumn{2}{|c|}{\textbf{Mean$\pm$Std}:} & \textbf{0.0036$\pm$0.0005} & \textbf{0.461$\pm$0.045}\\
\hline
\end{tabular}
\end{table}

\item External cross-talk.

SiPMs are known to emit infrared light when an avalanche is triggered~\cite{cross_ir}. The light could trigger nearby micro-cells of the same SiPM, resulting in what is known as optical cross-talk~\cite{cross}. When measuring reflectivity of a SiPM, the effect could in principle lead to the infrared light reaching the specular and diffuse background detectors, systematically affecting the reflectivity. A somewhat similar effect was described in Ref.~\cite{cross_fbk}. We check for such external cross-talk by performing the reflectivity measurement with a SiPM in the reflector position turned off and biased above the breakdown voltage (up to 4 V over-voltage). No difference in the apparent reflectivity is observed. 

\item Secondary reflections from specular detector.

Presence of a specular detector during PDE measurements increases the apparent light flux in the reflector, because a fraction of the photons hitting the specular detector will reflect back and hit the reflector again. Once the reflectivity of the SiPMs in the reflector and specular detector positions is known, the amount of the effect can be calculated using MC. For the detectors used in this work, the effect is found to be $\sim$2-5\%, depending on the angle of incidence. For precise measurement of the PDE, we do not include a specular detector in the setup. In such a case, the majority of the photons reflected from the reflector pass the opening in the copper rail and get absorbed by the coated copper cup (Figure~\ref{fig:lixo}), which eliminates the effect. 
\end{itemize}

\section{Validation of the approach}
To validate the measurement approach we perform reflectivity measurement with a sample that has independently known expected reflectance in the VUV. The sample is a piece of silicon wafer with a 1.5 $\si{\micro}$m silicon oxide layer, which was provided to the nEXO collaboration by FBK. The sample was measured by IHEP, Beijing, in vacuum at different VUV wavelengths and angles of incidence, allowing one to constrain the refractive indices of silicone oxide~\cite{ihep}. Fresnel equations are then used to calculate the expected reflectance of the xenon scintillation light in LXe. The reflectance value of 52.2$\pm$1.6\% is expected at 15$^{\circ}$ of incidence. As the incident angle increases, the reflectance of the wafer in LXe is expected to increase. We do observe an increase, but we can't compare it to the expectation, because the amount of the increase at larger angles of incidence begins to depend strongly on the (so far unconstrained) value of the LXe's index of refraction. So the validation test is restricted to small incident angles (15$^{\circ}$).

Table~\ref{table:val} summarizes the results of the validation measurement. During the measurement, the sample is placed in the reflector position using a PEEK holder, similarly to other samples. $R$ in the table denotes specular reflectance measured by LIXO, while $R_{\textup{exp}}$ denotes the value expected based on the vacuum measurement by IHEP, Beijing. The measurement of the wafer is in good agreement with the expectation, which suggests that the LIXO setup can be used to make accurate measurements and that the systematic uncertainties are correctly accounted for. 
\begin{table}[h]
\centering
\caption{Validation measurement. Reflectivity of a silicon wafer sample measured at 15$^{\circ}$ of incidence in LIXO, $R$, is compared with the expected value, $R_{\textup{exp}}$, based on the vacuum measurement by IHEP, Beijing.}
\label{table:val}
\begin{tabular}{|c|c|}
\hline
Quantity & Si wafer\\
\hline
\hline
$\mu^\textup{S}$, p.e. & 0.0467(4) \\
$\mu^\textup{R(0)}$, p.e. & 0.1022(4) \\
$\mu^\textup{D}$, p.e. & 0.0036(5) \\
Sys. correction, rel.\% & 16.2$\pm$3.2\\
Light stability error, rel.\% & 3 \\
\hline
\hline
\textbf{$R$, \%} & \textbf{50.8$\pm$2.3} \\
\hline
\textbf{$R_{\textup{exp}}$, \%} & \textbf{52.2$\pm$1.6} \\
\hline
\end{tabular}
\end{table}

By default, we operate the VUV4 SiPM devices at 3 V over-voltage. Higher over-voltages do not provide substantial increase in PDE, but increase the rate of dark counts and correlated avalanches. This may complicate the analysis by increasing the coincidence rates between the triggering collimator detector and other detectors. Nevertheless, we repeat the validation measurement at 4 V over-voltage as a cross-check. 
Table~\ref{table:val4} summarizes the results. Since the measurements at the two over-voltages were performed during the same liquefaction, the light stability uncertainty and the uncertainty of the systematic correction cancel out. $R_{4V}$ in the table denotes specular reflectance of the wafer measured with the specular detector operated at 4 V over-voltage.
\begin{table}[h]
\centering
\caption{Cross-check at different over-voltage. By default, detectors are operated at 3 V over-voltage. Additional measurement is performed at 4 V over-voltage, R$_{\textup{4V}}$, and the results are compared.}
\label{table:val4}
\begin{tabular}{|c|c|}
\hline
Quantity & Si wafer\\
\hline
\hline
$\mu^\textup{S}_{\textup{4V}}$, p.e. & 0.0508(4)\\
$\mu^\textup{R(0)}_{\textup{4V}}$, p.e. & 0.1086(4) \\
$\mu^\textup{D}_{\textup{4V}}$, p.e. & 0.0044(7) \\
Relative error, rel.\% & 1.2 \\
\hline
\hline
\textbf{$(R_{\textup{4V}}-R)/\sigma$} & \textbf{1.5}\\
\hline
\end{tabular}
\end{table}

\section{Results}

The methods described in Section~\ref{sec:meas} are used to measure relative PDE and reflectivity of Hamamatsu VUV4 SiPMs for xenon scintillation light in LXe. The results of the measurements are summarized below. 

\subsection{PDE measurement}
\begin{figure}[ht]
 \centering
\includegraphics[width=0.7\textwidth]{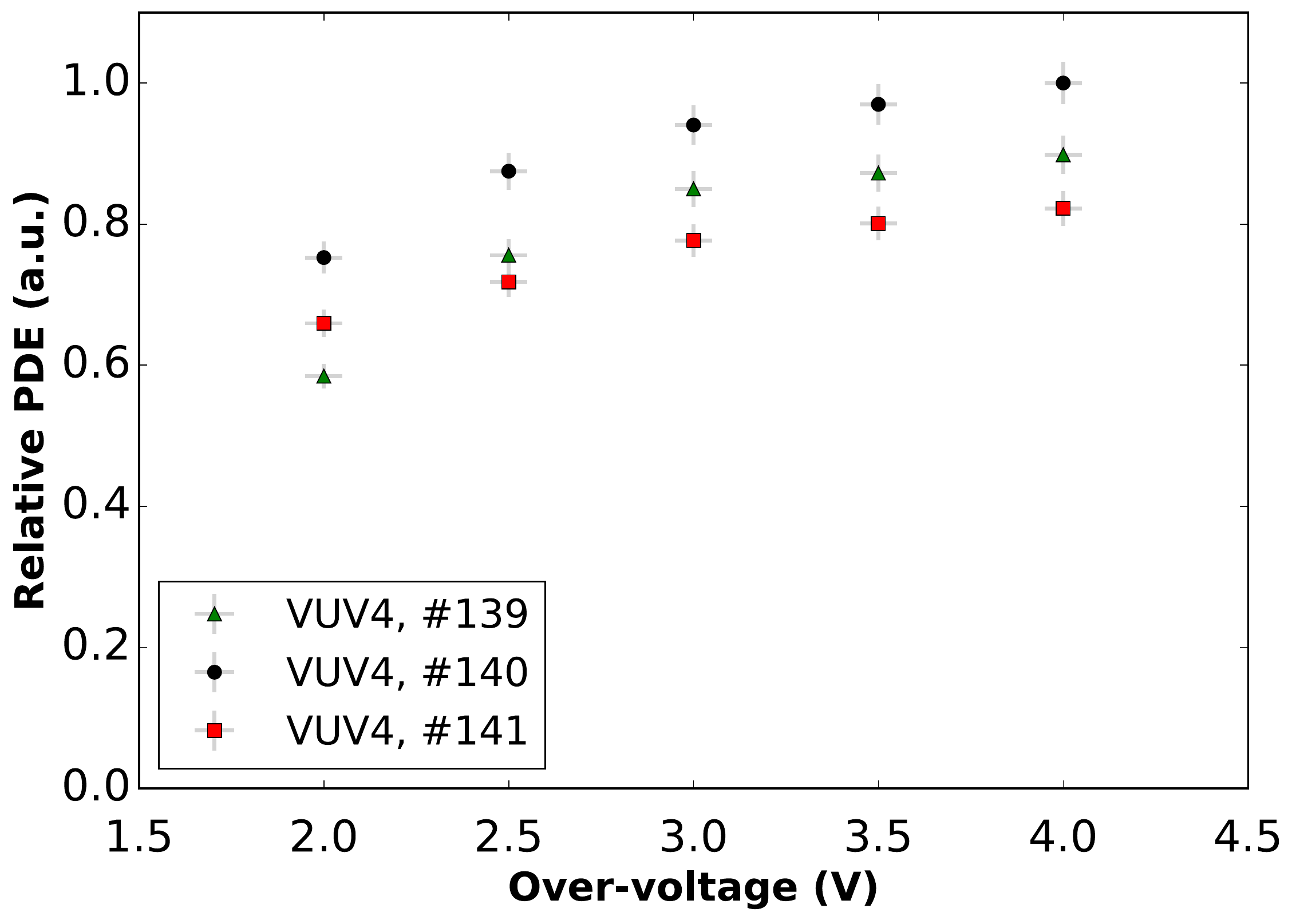}
  \caption{Relative PDE in LXe of three 50 $\si{\micro}$m micro-pixel VUV4 SiPM with serial numbers 139 (green triangles), 140 (black circles), and 141 (red squares).}
 \label{fig:pdeov}
\end{figure}

Figure~\ref{fig:pdeov} shows the measured relative PDE as a function of over-voltage for the three 50 $\si{\micro}$m micro-pixel VUV4 SiPMs with serial numbers 139 (green triangles), 140 (black circles), and 141 (red squares).
The measurements were performed at 0$^\circ$ of incidence. The Y-axis units are chosen such that the maximum observed PDE has the value of 1. The Y-axis error bars correspond to the sum (in quadrature) of the light level stability, the diffuse background, and the statistical errors. The PDE differs by roughly $\pm$8\% (Std) among the three devices. 

\begin{figure}[h]
 \centering
\includegraphics[width=0.7\textwidth]{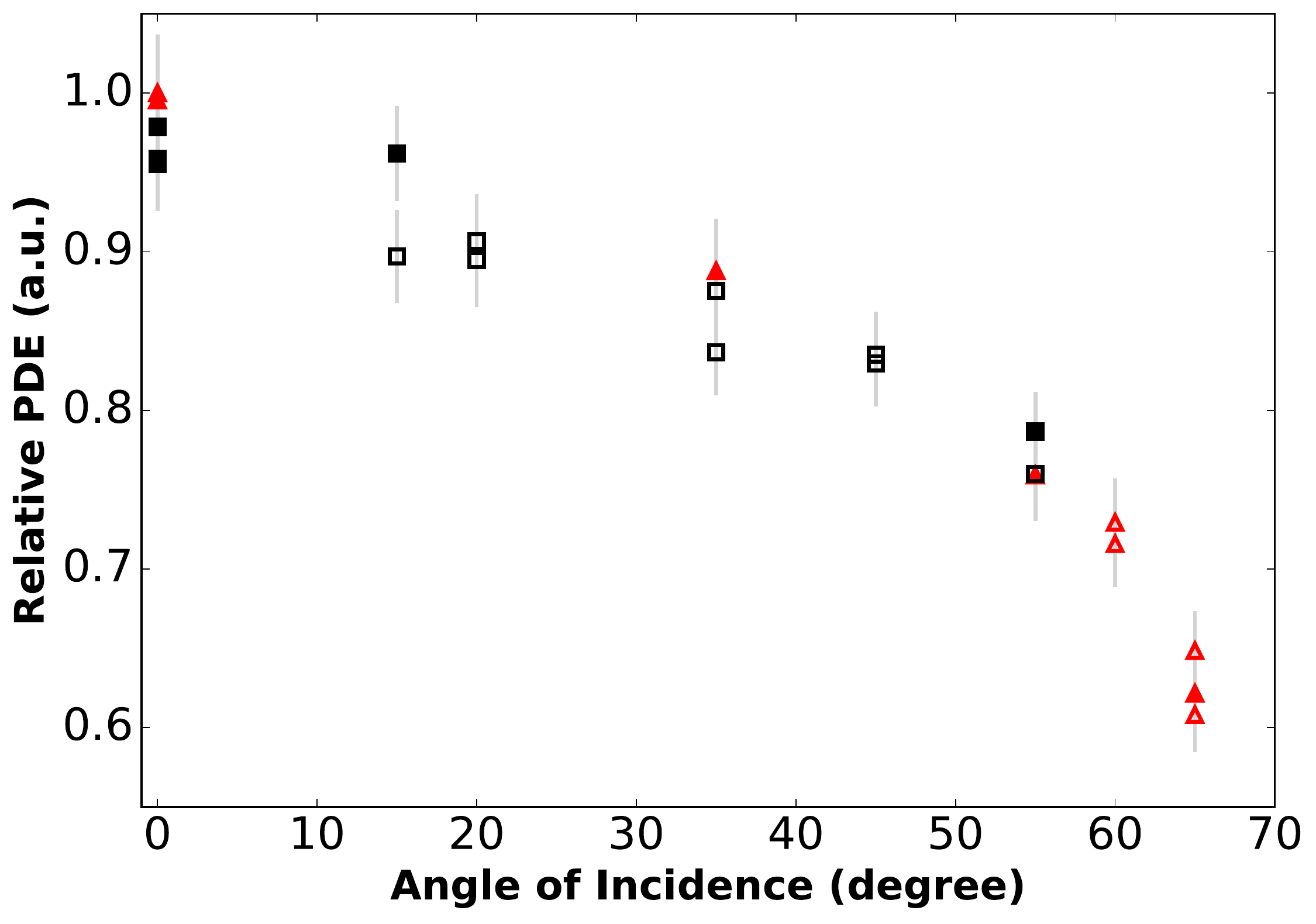}
  \caption{Relative PDE of the VUV4 SiPM serial number 141 in LXe as a function of the angle of incidence. Measurements were performed with two sources, denoted by black square and red triangle markers.}
 \label{fig:pdeang}
\end{figure}

For one of the SiPMs (serial number 141), the PDE was measured at 3 V over-voltage as a function of the angle of incidence  (Figure~\ref{fig:pdeang}). Two sources, marked by black squares and red triangles in the figure, were used to measure at two different angles during one liquefaction. Data points denoted by open markers correspond to measurements in which a specular detector was also present. Results of such measurements include an additional correction, as described in Section~\ref{sec:sys_summary}. The data points denoted by solid markers correspond to measurements in which no specular detector was present. The good agreement between the data points with open and solid markers suggests that the MC can be trusted to accurately estimate the correction. More than one measurement was performed for each angle of incidence in different liquefactions, separated by days or weeks, to check reproducibility. The Y-axis units are chosen such that the maximum observed PDE has the value of 1. The Y-axis error bars correspond to the sum (in quadrature) of the light level stability, the diffuse background, and the statistical errors. The PDE decreases as the angle of incidence increases. At 65$^{\circ}$ incidence the observed PDE is ca. 60\% of that at normal incidence.

\subsection{Reflectivity measurement}
Table~\ref{table:ref} summarizes the results of the specular reflectivity measurements for the three 50 $\si{\micro}$m VUV4 SiPM devices.
\begin{table}[h]
\centering
\caption{Reflectivity results at 15$^{\circ}$ for the three VUV4 SiPMs.}
\label{table:ref}
\begin{tabular}{|c|c|c|c|}
\hline
Quantity & \#139 & \#140 & \#141\\
\hline
$\mu^\textup{S}$, p.e. & 0.0273(3) & 0.0259(2) & 0.0274(3) \\
$\mu^\textup{R(0)}$, p.e. & 0.0943(9) & 0.0943(9) & 0.1022(4) \\
$\mu^\textup{D}$, p.e. & \multicolumn{3}{|c|}{0.0036(5)}  \\
Sys. correction, rel.\% & \multicolumn{3}{|c|}{16.2$\pm$3.2} \\
Light stability error, rel.\% & \multicolumn{3}{|c|}{3} \\
\hline
\hline
\textbf{$R$, \%} & \textbf{30.4$\pm$1.4} & \textbf{28.6$\pm$1.3} &  \textbf{28.0$\pm$1.3}\\
\hline
\end{tabular}
\end{table}
The measurements were conducted at 15$^{\circ}$ incident angle. The detectors were biased at 3 V over-voltage. The alpha cut on the signal in the triggering collimator detector was used in the analysis. The error is calculated by adding (in quadrature) the light stability error, the error on the systematic correction, and the diffuse background uncertainty. 

\begin{figure}[h]
 \centering
\includegraphics[width=0.7\textwidth]{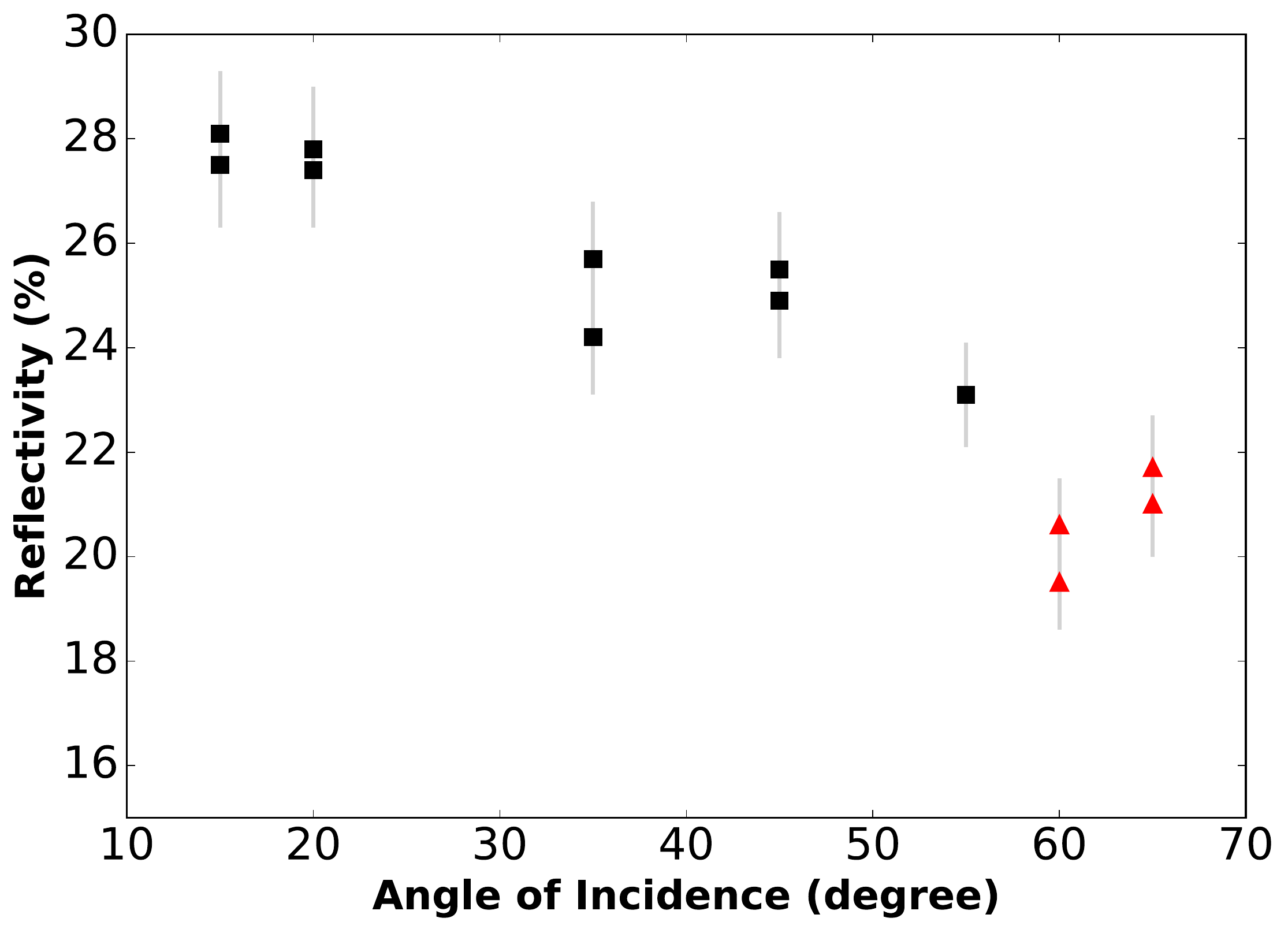}
  \caption{Specular reflectivity of the VUV4 SiPM serial number 141 in LXe as a function of the angle of incidence. Measurements were performed with two sources, denoted by black square and red triangle markers.}
 \label{fig:refang}
\end{figure}

For one of the SiPMs (serial number 141), the specular reflectivity was measured as a function of the angle of incidence  (Figure~\ref{fig:refang}). The reflectivity is found to decrease with the angle of incidence. 

\section{Conclusions}
Reflectivity and PDE of photosensors in LXe is an important input to the nEXO detector design. These parameters affect the light collection, and thus impact the energy resolution and sensitivity. While a lot of characterization can be done in vacuum, the ultimate comparison is best performed in LXe. 

This work investigates specular reflectivity and relative PDE of the VUV4 Hamamatsu SiPM (series:13370) in LXe. We find the three tested 50 $\si{\micro}$m micro-pixel devices to have the specular reflectivity at 15$^{\circ}$ of incidence of approximately 30\%, consistent among the devices. 
The PDE of the three SiPMs differs by about $\pm$8\% (Std) among the devices. A comparable difference in PDE of different VUV4 devices of the same series has been previously observed in Ref.~\cite{giacomo}. An angular scan of the PDE was performed for one of the devices. The PDE is found to decrease with the angle of incidence, with the detector being approximately 60\% as efficient in detecting the xenon scintillation light in LXe at 65$^{\circ}$, compared to the normal incidence. Both the PDE and the specular reflectivity are found to decrease with the angle of incidence. One possible cause is an imperfect transparency in the VUV of the top layer above the silicon. 

A special sample with independently known expected reflectivity values in LXe has also been measured in LIXO. The reasonable agreement of the obtained results with the expectation indicate that the LIXO setup can be used to make accurate reflectivity measurements and that the systematic uncertainties are correctly accounted for.

The preliminary studies conducted by nEXO to estimate the light collection efficiency and its impact on the energy resolution and sensitivity assumed no angular dependence of either reflectivity or PDE of the SiPMs or passive materials~\cite{nEXO}. This work provides the first input that will allow one to refine the preliminary estimates by incorporating the actual reflectivity and PDE data obtained in LIXO. 

Future work will include measurements of other photosensors and passive materials in LXe. Improvements to the LIXO setup are also foreseen. The two main goals are to allow one to continuously change the incident and reflected angles during one LXe run and to understand and suppress the diffuse background. 

\acknowledgments
This work is supported by a Department of Energy Grant No. DE-SC0019261. We gratefully acknowledge the support of Nvidia Corporation with the donation of the Titan Xp GPUs used for optical simulations conducted in this work. Support for nEXO comes from the Office of Nuclear Physics within DOE's Office of Science, and NSF in the United States, from NSERC, CFI, FRQNT, NRC, and the McDonald Institute (CFREF) in Canada, from IBS in Korea, from RFBR (18-02-00550) in Russia, and from CAS and NSFC in China.

\bibliographystyle{JHEP} 
\bibliography{main}

\end{document}